\documentclass[fleqn,usenatbib]{mnras}

\usepackage{newtxtext,newtxmath}
\usepackage[T1]{fontenc}
\usepackage{graphicx}	
\usepackage{amsmath}	
\usepackage{float}
\usepackage[style=base]{caption}
\usepackage{multirow}
\usepackage{hyperref}
\usepackage{natbib}
\usepackage[usenames]{color}
\graphicspath{{figs/}}
\usepackage[dvipsnames]{xcolor}
\usepackage{array}
\usepackage{footmisc}
\usepackage{balance}
\usepackage{lipsum}
\usepackage{ragged2e}

\newcolumntype{C}[1]{>{\centering\arraybackslash}m{#1}}
\newcolumntype{L}[1]{>{\arraybackslash}m{#1}}

\newcommand{\sref}[1]{Section~\ref{#1}}

\newcommand{\fref}[1]{Figure~\ref{#1}}

\newcommand{\tref}[1]{Table~\ref{#1}}

\newcommand{\kms}{\ifmmode  \,\rm km\,s^{-1} \else $\,\rm km\,s^{-1}$ \fi }
\newcommand{\Msun}{\ifmmode {M_{\odot}} \else ${M_{\odot}}$ \fi}

\def\zls{z_{\ell,\rm spec}}

\def\zlp{z_{\ell,\rm phot}}

\usepackage[usenames]{color}

\title[Strong Lens Finding in The HSC-SSP using CNN]{Survey~of~Gravitationally~lensed~Objects~in~HSC~Imaging~(SuGOHI) $-$ X. Strong~Lens~Finding~in~The~HSC-SSP~using~Convolutional~Neural~Networks}
\author[A. T. Jaelani et al.]{Anton T. Jaelani$^{1,2\thanks{E-mail: \href{antontj@itb.ac.id}{antontj@itb.ac.id}}}$,
Anupreeta More$^{3,4}$,
Kenneth C. Wong$^{5,6}$,
Kaiki T. Inoue$^{7}$,
Dani C. -Y. Chao$^{8}$,\newauthor 
Premana W. Premadi$^{1}$, and
Raoul Ca\~nameras$^{9,10}$
\medskip\\
$^1$Astronomy Research Group and Bosscha Observatory, FMIPA, Institut Teknologi Bandung, Jl. Ganesha 10, Bandung 40132, Indonesia\\
$^2$U-CoE AI-VLB, Institut Teknologi Bandung, Jl. Ganesha 10, Bandung 40132, Indonesia\\
$^3$The Inter-University Centre for Astronomy and Astrophysics (IUCAA), Post Bag 4, Ganeshkhind, Pune 411007, India\\
$^4$Kavli Institute for the Physics and Mathematics of the Universe (WPI), UTIAS, The University of Tokyo, Kashiwa, Chiba 277-8583, Japan\\
$^5$Research Center for the Early Universe, Graduate School of Science, The University of Tokyo, 7-3-1 Hongo, Bunkyo-ku, Tokyo 113-0033, Japan\\
$^6$National Astronomical Observatory of Japan, 2-21-1 Osawa, Mitaka, Tokyo 181-8588, Japan\\
$^7$Department of Physics, Kindai University, 3-4-1 Kowakae, Higashi-Osaka, Osaka 577-8502, Japan\\
$^8$Ru\dj er B\v{o}skovi\'{c} Institute, Bijeni\v{c}ka cesta 54, 10000 Zagreb, Croatia\\
$^9$Max-Planck-Institut f\"{u}r Astrophysik, Karl-Schwarzschild-Str. 1, D-85748 Garching, Germany\\
$^{10}$Technical University of Munich, TUM School of Natural Sciences, Department of Physics, James-Franck-Str. 1, D-85748 Garching, Germany
}
\date{Accepted XXX. Received YYY; in original form ZZZ}

\pubyear{2022}

\begin{document}
\label{firstpage}
\pagerange{\pageref{firstpage}--\pageref{lastpage}}
\maketitle

\begin{abstract}
We apply a novel model based on convolutional neural networks (CNNs) to identify gravitationally-lensed galaxies in multi-band imaging of the Hyper Suprime Cam Subaru Strategic Program (HSC-SSP) Survey. The trained model is applied to a parent sample of 2~350~061 galaxies selected from the $\sim$ 800 deg$^2$ Wide area of the HSC-SSP Public Data Release 2. The galaxies in HSC Wide are selected based on stringent pre-selection criteria, such as multiband magnitudes, stellar mass, star formation rate, extendedness limit, photometric redshift range, etc. Initially, the CNNs provide a total of 20~241 cutouts with a score greater than 0.9, but this number is subsequently reduced to 1~522 cutouts by removing definite non-lenses for further inspection by human eyes. We discover 43 definite and 269 probable lenses, of which 97 are completely new. In addition, out of 880 potential lenses, we recovered 289 known systems in the literature. We identify 143 candidates from the known systems that had higher confidence in previous searches. Our model can also recover 285 candidate galaxy-scale lenses from the Survey of Gravitationally lensed Objects in HSC Imaging (SuGOHI), where a single foreground galaxy acts as the deflector. Even though group-scale and cluster-scale lens systems were not included in the training, a sample of 32 SuGOHI-c (i.e., group/cluster-scale systems) lens candidates was retrieved. Our discoveries will be useful for ongoing and planned spectroscopic surveys, such as the Subaru Prime Focus Spectrograph project, to measure lens and source redshifts in order to enable detailed lens modelling.
\end{abstract}



\section{Introduction}
\label{sec:intro}
As strong gravitational lensing systems are discovered over large cosmological volume, they are powerful astrophysical tools to refine cosmological models and study the evolution of cosmic large scale structures. Studies of strong lens systems have been successfully used to probe the total mass distribution (including both baryonic and dark matter) from galaxy \citep[e.g.,][]{koopmans+06,auger+10,Son+15,shajib+2021} to group and cluster scales \citep[e.g,][]{limousin+09,More12,oguri+12,newman+13,jauzac+2021,allingham+2023}, constraints on dark-matter substructures \citep[e.g.,][]{vegetti+10,vegetti+12,hezaveh+16,inoue+16,nierenberg+2017,gilman+2020,hsueh+2020,kaiki23}, and the lensing magnification effect can be exploited to study compact high-redshift objects in detail by overcoming the sensitivity and/or resolution limits of current facilities \citep[e.g.,][]{marques+17,more+17,Jae20}. Moreover, strongly-lensed quasars with time delay measurements can be used to test the cosmological model by constraining cosmological parameters such as the Hubble constant \citep[e.g.,][]{refsdal64,wong+20b,birrer+20}.

\begin{table*}
	\centering
	\caption{Summary of previous SuGOHI lens searches.}
	\label{tab:table1}
	\begin{tabular}{C{1.1cm}C{1.5cm}C{1.5cm}C{4.1cm}C{2.5cm}C{4.5cm}} 
		\hline
		SuGOHI & Data Release & Area (deg$^2$) & Inspected objects & Grades (A, B, C) & Description \\
		\hline
		 I    & up to S16A & 442  & $\sim$ 43~000 (LRG)                      & (15,  36, 282) & {\sc YattaLens}$^{\dagger}$, {\sc Chitah}$^+$, Spectroscopic \\
	     II   & up to S17A & 776  & 31~286  (LRG)                           & (7,  34, ...)  & {\sc YattaLens}$^{\dagger}$ \\
		 IV   & up to S16A & 456  & $\sim$ 45~000 (LRG); $\sim$ 110,000 (QSO) & (17,  26, ...) & {\sc Chitah}$^+$ \\
		 V    & up to S18A & 1~114 & $\sim$ 39~500 (Galaxy Clusters)          & (47, 181, 413) & Cluster catalogues, Human-inspection\\ 
		 VI   & up to S16A & 442  & $\sim$ 300~000                           & (14, 129, 581) & Citizen-Scientist \\
		 VII  &   ...      & ...  & SQLS; Chandra catalogue            & (3, ..., ...)  & reinspection previous catalogues\\
		 VIII & up to S21A & 1310 & 103~191 (LRG)                           & (8,  28, 138)  & {\sc YattaLens}$^{\dagger}$ \\
         IX    & up to S21A & 1~310 &  1~652~329 (Quasar) & (73, 17, 53) & colour similarity, {\sc YattaLens}$^{\dagger}$ \\ 
		\hline
	\end{tabular}
	{\justifying \textbf{Notes}. $^{\dagger}$I \citep{Son18}; II \citep{Wong18}; IV \citep{Chan20}; V \citep{Jae20}; VI \citep{Son20}; VII \citep{Jae21}; VIII \citep{Wong+22}; IX \citep{Chan+23}; $^+$\citep{Chan+15}. LRG is Luminous Red Galaxies, QSO is Quasi-Stellar Objects, and SQLS is Sloan Digital Sky Survey Quasar Lens Search.
	\par}
\end{table*}

In the last decade, deep learning has led to very good performance on a variety of problems, such as visual recognition, speech recognition and natural language processing, including the application in astrophysical problems e.g, strong gravitational lensing detection. Among different types of deep neural networks, convolutional neural networks (CNNs) have been most extensively studied. \cite{Lecun+98} published the seminal paper establishing the improved modern framework of CNNs. The CNNs have proven extremely efficient as 'state of the art' for pattern recognition tasks and have given a strong impetus to image analysis and processing. Recent studies largely demonstrate the ability of supervised CNNs to identify the rare gravitational lenses among large datasets \citep[e.g.,][]{Jacobs+17,Jac19,Pet19,Can20,Huang21,Li+21,Zaborowski23}, extending previous automated algorithms \citep[e.g.,][]{Gav14} generally with better classification performance \citep{metcalf19}. 

The Survey of Gravitationally lensed Objects in HSC Imaging (SuGOHI) is an ongoing lens search that aims to discover a large number of strong gravitational lens systems from the large homogeneous data of the Hyper Suprime-Cam Subaru Strategic Program (HSC-SSP).  SuGOHI has discovered about 2300 strong gravitational lens candidates at both the galaxy-scale \citep[SuGOHI-g;][]{Son18,Son20,Wong18,Wong+22}, and the galaxy group/cluster-scale \citep[SuGOHI-c;][]{Jae20}, including a number of lensed quasar candidates \citep[SuGOHI-q;][]{Chan20,Jae21,Chan+23}. These lenses were discovered using a variety of search techniques, including semi-automated algorithms, visual inspection, and citizen science (see summary in \tref{tab:table1}). In this study, we apply CNNs to identify new strong gravitational lens systems. \citet{Can21} and \citet{Shu+22} have independently applied machine learning-based lens search algorithms with different networks to the HSC SSP Public Data Release 2 (PDR2) \citep{Aihara+19} and published lens candidates produced by these searches. 

The paper is organised as follows. First, in \sref{sec:hsc} we provide a brief overview of the HSC-SSP PDR2 and the definition of our parent sample. In \sref{sec:sims}, the system of simulated lenses is described. In \sref{sec:cnn}, We discuss the networks, training procedures, CNN's performance, and visual inspection. \sref{sec:res} presents new lens candidates. Discussions and conclusions are presented in \sref{sec:summ}. The AB system is used for all magnitudes. All images are oriented with North up and East to the left.

\section{The HSC-SSP PDR2}
\label{sec:hsc}

\begin{figure*}
\includegraphics[width=\textwidth]{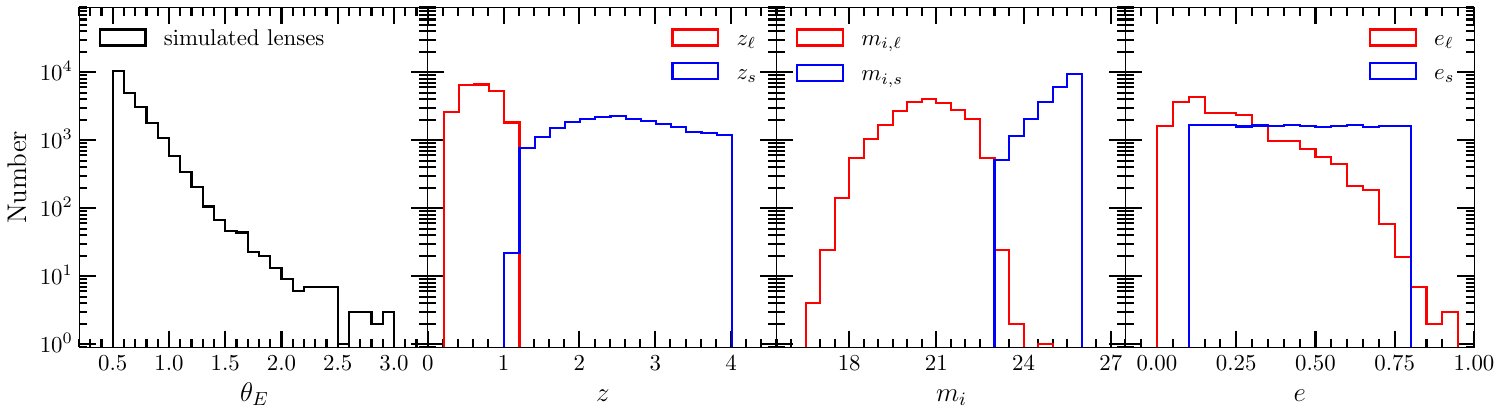}
\caption{The distribution of the properties of the simulated lenses is constructed by {\sc simct}. Left to right: the Einstein radius, $\theta_E$; the redshift, $z$; the {i}-band magnitude, $m_i$; and the ellipticity, $e$. The red and blue colours denote lens and source quantities, respectively.}
\label{fig:mock}
\end{figure*}

The HSC-SSP Survey is an optical imaging survey conducted with the Hyper Suprime-Cam \cite[HSC,][]{Miyazaki+18,Komiyama+18, Kawanomoto+18, Furusawa+18,Huang+18,Coupon+18}, a $1.7\deg^2$ field-of-view optical camera on the 8.2-meter \textit{Subaru telescope}, which has a pixel scale of 0.168\arcsec/pixel. In this work, we use data products from the HSC-SSP Public Data Release 2 (henceforth ``PDR2") Wide covering $\sim$ 800 deg$^2$ \citep{Aihara+19} in \textit{gri} bands out of a final survey area of 1400 deg$^2$. The median $5\sigma$ depths (for point sources) in $gri$ filters for the PDR2 Wide layer are 26.6, 26.2, and 26.2 mag, while the median seeings are 0.77\arcsec, 0.76\arcsec, and 0.58\arcsec, for $gri$, respectively. The PDR2 data is processed with \texttt{hscPipe} version 6 \citep{Bosch+18}. The combination of a large area and observation depth makes the data from the HSC-SSP Survey ideal for discovering new or unusual strong gravitational lenses \citep[e.g., HSC J090429.75-010228.2;][]{Jae20b}.

The photometric redshifts used in this work are determined using the \texttt{mizuki} algorithm \citep[][]{Tanaka+15}. The robustness of the photometric redshifts is a function of galaxy redshift and brightness, and is quantified in terms of $\Delta z/(1 + z_{\rm ref})$, where ${\rm{\Delta }}z\equiv | z-{z}_{\rm ref}| $ and $z_{\rm ref}$ is a reference redshift. The \texttt{mizuki}'s photometric redshifts measurement are most accurate at $0.2 \lesssim z_{\rm phot}\lesssim 1.5$. A thorough explanation of \texttt{mizuki}'s application to the HSC-SSP data is presented in \cite{Tanaka+18}. Since the HSC-SSP Survey footprint overlaps with the Sloan Digital Sky Survey (SDSS), we collect spectroscopic redshifts, whenever available, from the SDSS Data Release 16 \citep{Ahumada+20} catalogues.

For our strong lens search, we select to be our parent sample galaxies in PDR2 Wide retrieved from the HSC CAS Search service\footnote{https://hsc-release.mtk.nao.ac.jp/datasearch/} based on their magnitude limit, stellar mass, and star formation rate. To be more specific, we select objects from tables that satisfy the following criteria:
\begin{itemize}
    \item for \texttt{pdr2\_wide.forced}
    \begin{enumerate}
        \item[1.] \texttt{isprimary} is \texttt{True}
        \item[2.] \texttt{[grizy]\_pixelflags\_edge} is \texttt{False}
        \item[3.] \texttt{[grizy]\_pixelflags\_interpolatedcenter} is \texttt{False}
        \item[4.] \texttt{[grizy]\_pixelflags\_crcenter} is \texttt{False}
        \item[5.] \texttt{[grizy]\_cmodel\_flag} is \texttt{False}
        \item[6.] \texttt{i\_extendedness\_value} > 0.9
        \item[7.] \texttt{([ri]\_cmodel\_mag $-$ a\_[ri])} < 28.0
        \item[8.] \texttt{(z\_cmodel\_mag $-$ a\_z)} < 23.0
    \end{enumerate}

    \item for \texttt{pdr2\_wide.forced2}
    \begin{enumerate}
        \item[9.] \texttt{[grizy]\_sdsscentroid\_flag} is \texttt{False}
    \end{enumerate}
    
    \item \texttt{pdr2\_wide.meas}
    \begin{enumerate}
        \item[10.] \texttt{[grizy]\_inputcount\_value} > 0
    \end{enumerate}
    
    \item \texttt{pdr2\_wide.masks}
    \begin{enumerate}
        \item[11.] \texttt{i\_mask\_pdr2\_bright\_objectcenter} is \texttt{False}
    \end{enumerate}
    
    \item \texttt{pdr2\_wide.photoz\_mizuki}
    \begin{enumerate}
        \item[12.] 0.2 < \texttt{photoz\_median} < 1.2
        \item[13.] \texttt{stellar\_mass} > 5.0e10
        \item[14.] \texttt{(sfr / stellar\_mass)} < 1.0e-10.
    \end{enumerate}
\end{itemize}

We remove objects with saturated pixels and unreliable photometry, as well as probable stars, using criteria 2-11. We concentrate on finding lensed features among a set of galaxies based on the properties that make them possible lenses, as defined by criteria 12-14. The upper redshift and lower stellar mass bounds are a compromise between obtaining a sample of lenses as complete as possible and the need to find lenses with Einstein radius smaller than 1\arcsec. The lower redshift cut is introduced to avoid dealing with galaxies that are too bright for the detection of lensed features, which are typically faint. This query returns a total of 2~350~061 unique HSC objects. We use HSC $gri$ cutouts (64 pixels $\times$ 64 pixels, which is sufficient for galaxy-scale lenses) centred on the parent objects, which are retrieved from the PDR2 Image Cutout service\footnote{https://hsc-release.mtk.nao.ac.jp/das\_cutout/pdr2/}. 

\section{Simulated Galaxy-Scale Lenses}
\label{sec:sims}

\begin{figure}
\centering
\includegraphics[width=0.4\textwidth]{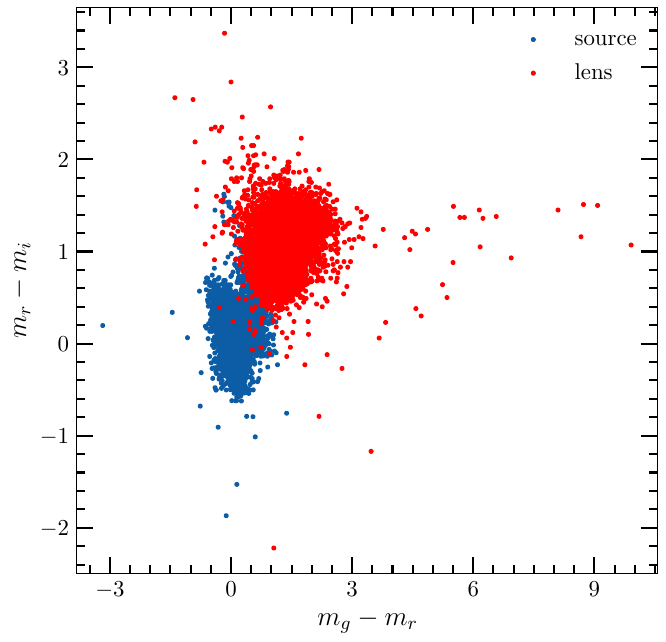}
\caption{Distributions of color-color ($m_r-m_i$ vs. $m_g-m_r$) diagram of the lens galaxies (red), and the sources (blue).}
\label{fig:mcolor}
\end{figure}

In order to train the network on a large sample of realistic galaxy-scale lenses, we use the {\sc simct}\footnote{\url{https://github.com/anupreeta27/SIMCT}} pipeline \citep[detailed in][]{More16}. With this pipeline, we superpose a set of simulated lensed galaxies on the real image cutouts of massive galaxies from  the multi-band HSC imaging (see \sref{sec:hsc}). This ensures that most of the features and properties in the training images will statistically be similar to the the real survey images which will be classified. The methodology of {\sc simct} is summarized below. 

The mass model of the lens galaxies is assumed to follow Singular Isothermal Ellipsoid \citep[SIE,][]{Kormann+94} and the contribution from an external shear is also included in the model (see \tref{tab:properties}). We start with the position, photometric magnitudes, photometric redshifts and ellipticity of the massive galaxies to convert their light parameters into the SIE model mass parameters. Thus, following the $L - \sigma_v$ scaling relation \citep{Parker+05}, we obtain the velocity dispersion. Similarly, assuming mass follows light, we obtain the position of the lens potential and the ellipticity parameters.

\begin{figure*}
\includegraphics[width=\textwidth]{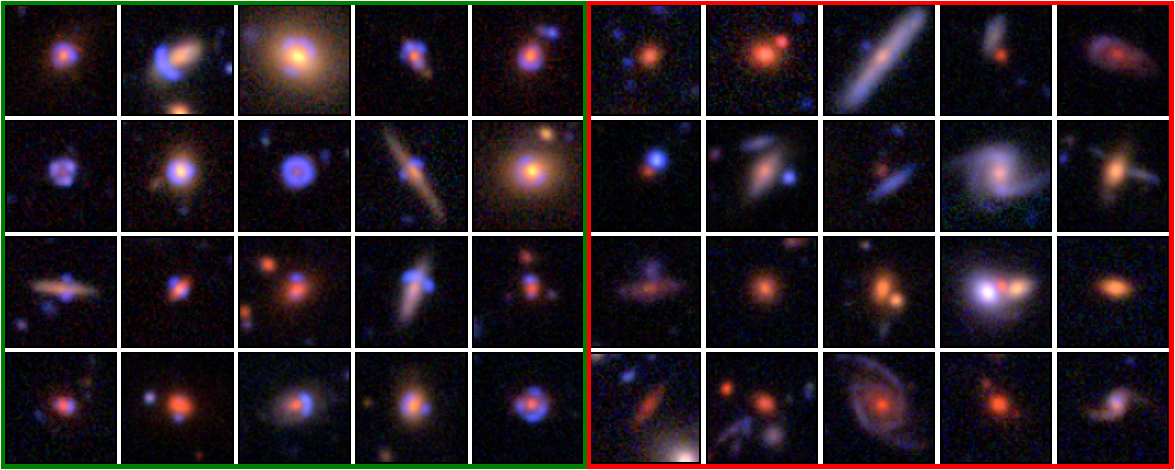}
\caption{\textit{Green box}: examples of simulated lenses based on real HSC SSP colour composite images. \textit{Red box}: examples of objects used as non-lenses during the training of the CNN. All postage stamps are 64 pixels on-a-side, corresponding to 10.8\arcsec.}
\label{fig:mock2}
\end{figure*}

\begin{table}
	\centering
	\caption{Thresholds used in the selection of the simulated lenses. }
	\label{tab:properties}
	\begin{tabular}{L{4cm}C{1.6cm}C{1.6cm}} 
		\hline
		Parameter & Range & Unit \\
		\hline
		\multicolumn{3}{c}{Lens (SIE)}\\
		\hline
		Einstein radius, $\theta_E$             & [0.5, 3.0]     & arcsec \\ 
		Redshift, $z_{\ell, \rm phot}$          & [0.2, 1.1]     & -      \\ 
		External Shear, $\gamma_{\rm ext}$      & [0.001, 0.020] &        \\
		PA of External Shear, $\theta_{\gamma_{\rm ext}}$ & [0, 180]       & degree \\
        \hline
		\multicolumn{3}{c}{Source}\\
		\hline
		Redshift, $z_s$                         & [1.2, 4.0]     & -     \\
		$i$-magnitude, $m_{i,s}$                & [23, 26]       & -     \\
		Ellipticity, $\theta_{e,s}$             & [0.1, 0.8]     & -      \\
		PA of Ellipticity, $\theta_{e,s}$       & [0, 180]       & degree\\
		\hline
	\end{tabular}
\end{table}

The $i$-band magnitude and the redshift of the background galaxies are drawn from the source counts and redshift distribution as prescribed in \citet{Fau09}. Faure et al. assumed that the number density of sources at redshift $z_s$ brighter than the apparent magnitude threshold of $m_{lim}\sim 26$ [mag] in the F814W filter band. Each galaxy is parameterised with a Sersic profile for a fixed Sersic index of 1 corresponding to an exponential profile. The ellipticity and position angle (PA) are drawn randomly from a uniform distribution within the range indicated in \tref{tab:properties}. The effective radius of the galaxy is estimated from the luminosity - size, $L - r_{\rm eff}$ relation \citep[][with a redshift scaling to account for size evolution]{Bernardi+03}. The colors of the galaxy are extracted from photometric CFHTLenS catalog \citep{Hildebrandt12, Erben13}.

We take into account the lensing optical depth to determine which massive galaxy is likely to act as a lens and accept a background source provided the Einstein radius falls within the range given in \tref{tab:properties} and the magnitude of the second brightest lensed image is above the limiting magnitude of the HSC Survey \citep[see][for further details]{More16}. Once all of the parameters for the lens and source models have been determined, we use {\sc gravlens} \citep{Keeton2001} to generate simulated lensed images. After accounting for the shot noise in the lensed images and convolving them with the seeing in each of the filters. The simulated lensed image is added to the real HSC image centred on the galaxy selected to act as a lens (see the left green box in \fref{fig:mock2}). \fref{fig:mock} shows the distributions of the Einstein radius, redshifts, $i$-band magnitude, and ellipticity for both lens and the source of the simulated galaxy-scale lenses. We also show the distribution of color-color ($m_r-m_i$ vs. $m_g-m_r$) diagram of the lens galaxies and the sources (see \fref{fig:mcolor}) with selection effect tend to bluer sources.

From a total of about 22~000 simulated lens systems, we exclude about 3000 systems located within GAMA09H field which will be used for a comparison analysis in More et al. (2023, in prep.). We also exclude tens of simulated lens galaxies whose centres have been identified as known lens galaxies in previous studies \citep[e.g.,][]{More12,Son18, Son20}. We then end up with
18~660 simulated lens systems.

\section{Convolutional Neural Network}
\label{sec:cnn}

The dataset for training the CNN is divided into two subsets of equal number of members: one containing the 18~660 simulated lenses simulated with 64 $\times$ 64 pixel (corresponding to 10.752\arcsec $\times$ 10.752\arcsec) cutouts from \sref{sec:sims}, and the other containing the 18 660 non-lens objects that were not used to generate simulated lenses. These cutouts are large enough to capture galaxy-scale lenses and yet small enough for fairly quick training and searching.

\begin{table}
	\centering
	\caption{Details of training and test samples used in the analyses.}
	\label{tab:samp}
	\begin{tabular}{L{2.1cm}L{4.3cm}C{0.8cm}} 
		\hline
		\multirow{2}{*}{Training Sample}  &  Lenses  & 14~928\\
		            &  Non-Lenses  & 14~928 \\
        \multirow{2}{*}{Validation Sample} &  Lenses & 3~732 \\
		            &  Non-Lenses  & 3~732  \\
        \multirow{2}{*}{Test Sample}     &  Lenses  & 220   \\
		            &  Non-Lenses  & 78~665 \\
        \hline
		\multicolumn{3}{c}{Composition of the Non-lens sample}\\
		\hline
		 \multirow{5}{*}{\parbox{2cm}{Non-Lenses used in Training}} & Random Galaxies      & 55\% \\
	    & Spiral Galaxies                  & 34\% \\
		& Stars (SDSS Spectroscopic star)  & 5\% \\
		& Groups or LRG + edge-on galaxies & 4\% \\
        & Doubles, 'tricky' spirals or mergers & 2\% \\
		\hline
	\end{tabular}
 {\justifying \textbf{Note:}See \sref{sec:cnn} for further details. 
 \par}
\end{table}

\begin{table}
	\centering
	\caption{CNN Architecture}
	\label{tab:layout}
	\begin{tabular}{C{0.35cm}C{1.80cm}C{0.9cm}C{0.5cm}C{1.25cm}C{1.25cm}} 
		\hline
		ID & Type & Size & $n$ & Activation & Param Count\\
		(1) & (2) & (3) & (4) & (5) & (6)\\
		\hline
		1  & input           & 64 $\times$ 64 & ... & ...     & ... \\
		2  & convolutional   & 11 $\times$ 11 &  64 & ReLU    & 23~296 \\
		3  & max pooling     &  2 $\times$ 2  & ... & ...     & ... \\
		4  & convolutional   &  7 $\times$ 7  & 128 & ReLU    & 401~536 \\   
        5  & dropout (0.2)   &     ...        & ... & ...     & ... \\      
        6  & convolutional   &  5 $\times$ 5  & 128 & ReLU    & 409~728 \\   
        7  & max pooling     &  2 $\times$ 2  & ... & ...     & ... \\        
        8  & convolutional   &  5 $\times$ 5  & 256 & ReLU    & 819~456 \\
        9  & dropout (0.2)   &     ...        & ... & ...     & ... \\     
        10 & convolutional   &  3 $\times$ 3  & 256 & ReLU    & 590~080 \\   
        11 & max pooling     &  2 $\times$ 2  & ... & ...     & ... \\    
        12 & fully-connected &  1 024          & ... & ReLU    & 2~360~320\\   
        13 & dropout (0.2)   &     ...        & ... & ...     & ... \\       
        14 & fully-connected &  1 024          & ... & ReLU    & 1~049~600\\  
        15 & dropout (0.2)   &     ...        & ... & ...     & ... \\ 
        16 & fully-connected &  512           & ... & ReLU    & 524~800\\  
        17 & dropout (0.2)   &     ...        & ... & ...     & ... \\   
        18 & fully-connected &  512           & ... & ReLU    & 262~656\\
        19 & fully-connected &  1             & ... & sigmoid & 513\\
		\hline
        \multicolumn{3}{l}{Total params:} & \multicolumn{3}{l}{6~441~985} \\
        \multicolumn{3}{l}{Trainable params:} & \multicolumn{3}{l}{6~441~985} \\
        \multicolumn{3}{l}{Nontrainable params:} & \multicolumn{3}{l}{0} \\
        \hline
	\end{tabular}
	{\justifying \textbf{Notes}. (1) ID of the layers. (2) Type of the layers. (3) Size of the data or the filters. (4) Number of the filters. (5) Activation function adopted in the layers. (6) Trainable parameters.
	\par}
\end{table}

The dataset is split into 80\% for training and 20\% for validation (see \tref{tab:samp}). We select non-lens objects for the negatives, which contains: 55\% galaxies that are randomly selected from the parent catalogue; 34\% spiral galaxies from \citet{Tadaki+20}; 5\% stars from the SDSS spectroscopic star catalogue \citep{Ahumada+20}; 4\% galaxy groups or "crowded" galaxies like LRG + egde-on galaxy (or arc like feature); and 2\% dual point-like, tricky spiral or merger galaxies that are visually inspected from the parent catalogue (see the right red box in \fref{fig:mock2}).

In order to identify blue sources, we select HSC $gri$-band images due to their excellent depth and quality. The FITS images in three bands are converted to RGB images scaled with an arcsinh stretch using {\sc HumVI}\footnote{https://github.com/drphilmarshall/HumVI} \citep{Marshall+16} and supplied to the CNN as vectors of 12~288 (64 $\times$ 64 $\times$ 3) floating-point numbers. Before the CNN's training stage, we preprocess our data by normalizing each image brightness to be between 0 and 1. In general, data augmentation is a common practise in machine learning; it is used to expand the training set in order to prevent overfitting and teach the network rotational, translational and scaling invariance. Here, we use in-place data augmentation to make sure that, when the network is trained, it sees new variations of our dataset at each and every epoch. This augmentation method is done at training time and not an "additive" operation which is transforming the original data in the batch by a series of random transformation, then returning the randomly transformed into network for training. We augment our dataset applying the following transformations to the simulated and non-lens galaxy sample: (i) a random rotation in the range [-30 deg, 30 deg]; (ii) a random resizing by a factor in the range [0.8, 1.2]; (iii) a random flip along the horizontal axis; (iv) and a random channel\_shift\_range $= 0.9$.

\begin{figure}
\includegraphics[width=0.48\textwidth]{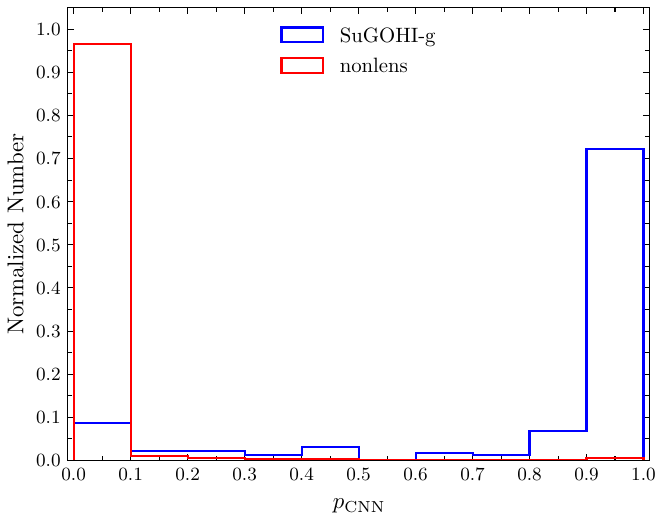}
\caption{Histogram of the probabilities predicted by the trained CNN for all lens and non-lens of test sample.}
\label{fig:performances2}
\end{figure}

\begin{figure}
\includegraphics[width=0.48\textwidth]{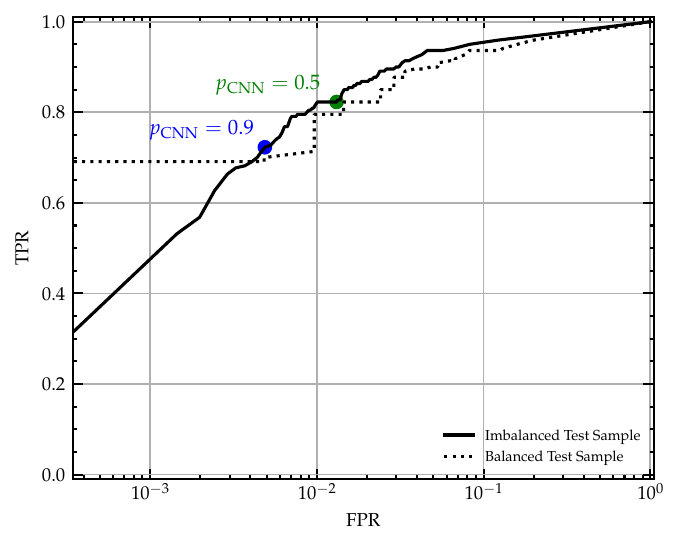}
\includegraphics[width=0.48\textwidth]{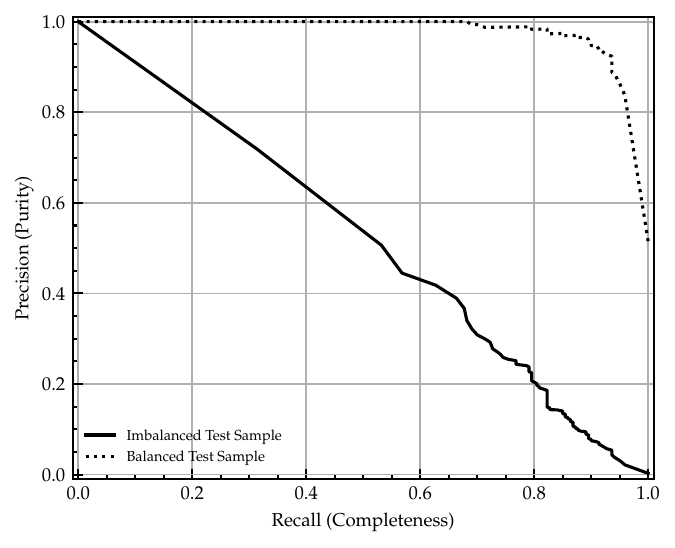}
\caption{\textbf{Top:} Receiver Operating Characteristic curves for the trained CNN showing the true position rate (TPR) as function of the false positive rate (FPR) for different lens identification thresholds. The corresponding area under curve (AUC) are 0.976 and 0.971, for imbalanced and balanced test sample, respectively. \textbf{Bottom:} Precision-Recall (PR) curves for the trained CNN showing the precision as function of the recall. Note that this performances describes a test dataset with a 50-50 split (for balanced) and a 0.3-99.7 split (for imbalanced) between lens and non-lens containing images. Number of test sample is shown in \tref{tab:result}.} 
\label{fig:performances}
\end{figure}

The CNN architecture is inspired by LeNet \citep{Lecun+98} and \citet{Jacobs+17}, with modifications displayed in \tref{tab:layout}. The CNN consists of five convolutional layers with kernel sizes of 11 $\times$ 11, 7 $\times$ 7, 5 $\times$ 5, 5 $\times$ 5, and 3 $\times$ 3; 64, 128, 128, 256, and 256 filters, followed by four fully connected hidden layers with 1~024, 1~024, 512, and 512 neurons. Five dropout regularizations \citep{Srivastava+14} are performed in between convolutional and fully-connected layers to minimise the likelihood of overfitting by dropping 20\% of the output neurons at random during training with Rectified Linear Unit \citep[ReLU;][]{Nair+10} nonlinear activations. The output layer consists of a single neuron with sigmoid activation. Three Maxpooling layers with $2 \times 2$ kernel sizes and stride 2 are inserted between the convolutional layers to make the CNN invariant to local translations of the relevant features in \textit{gri} image cutouts while reducing the network parameters. The network has a total of 6 441 985 trainable parameters upon completion. The network is summarised in \tref{tab:layout}.

We use the Adam optimization algorithm \citep{kingma+14} to minimise the cross-entropy error function over the training data with a learning rate of 0.00005. The CNN is trained for 100 epochs using mini-batch stochastic gradient descent with 128 images per batch. If the network does not provide improved accuracy or loss after five consecutive epochs, we terminate the training early. This network is implemented utilising the Python neural network library Keras \citep{Chollet+18} with the TensorFlow backend \citep{Abadi+16}.

Completeness is measured on spectroscopically-confirmed or grade A or B SuGOHI galaxy-scale lenses \citep{Son18,Wong18,Son20}. We rejected a few lenses with large image separations $\gtrsim3\arcsec$, indicating significant perturbation from the lens environment, as we have no intention of recovering such configurations. As test lenses, we use 220 SuGOHI-g lens systems. We collect the 78~665 non-lenses, including random galaxies from a parent catalogue; spiral galaxies from \citet{Tadaki+20}; and  stars from the SDSS spectroscopic star catalogue \citep{Ahumada+20}.

\subsection{CNN Performance}
\label{sec:cnnperf}

The CNN provides a score, $p_{\rm CNN}$, ranging from 0 to 1, with 1 for lenses and 0 for everything else (non-lenses). The evolution of the training and validation loss for the network with optimised hyperparameters until epoch 52, which corresponds to the validation set's minimal binary cross-entropy loss without overfitting. The network achieved a training (validation) accuracy of 0.986 (0.981) and a loss of 0.041 (0.059). The small difference between the two curves (generalisation gap) indicates that the model's predictions on new data with similar properties to the training set are relatively stable. Using the test set, the final network performance was characterised. In \fref{fig:performances2}, we show the probabilities predicted by the model for all lens and non-lens test examples. When $p_{\rm CNN}>0.9$, the distribution is dominated by lenses. 

To evaluate the number of lenses correctly identified, we use a Receiver Operating Characteristic (ROC) curve  which compares the True Positive Rate (TPR) to the False Positive Rate (FPR) as functions of the decision threshold applied to the score and Precision-Recall (PR) curves for the trained CNN showing the precision as function of the recall or TPR. In general, the excellent performance are shown by the ROC and PR curves in \fref{fig:performances}. The ROC curve illustrates the performance of a binary classifier in distinguishing between the two classes as the decision threshold is altered. The Area Under the ROC Curve (AUC) of our CNN are 0.976 and 0.971, for imbalanced (0.3-99.7) and balanced (50-50) test sample, respectively. For balanced test sample, the PR curve shows that the network can classifies all objects as lens sample would yield a precision (purity) of 97.3\% at a recall (completeness) of 82.3\% of SuGOHI-g lenses in the test set with an FPR of 1.3\% for $p_{\rm CNN}=0.5$ as the threshold; and a precision of 98.8\% at a recall of 72.3\% of SuGOHI-g lenses with an FPR of 0.49\% for $p_{\rm CNN}=0.9$ as a threshold (see \tref{tab:result}). 

\begin{figure}
\includegraphics[width=0.48\textwidth]{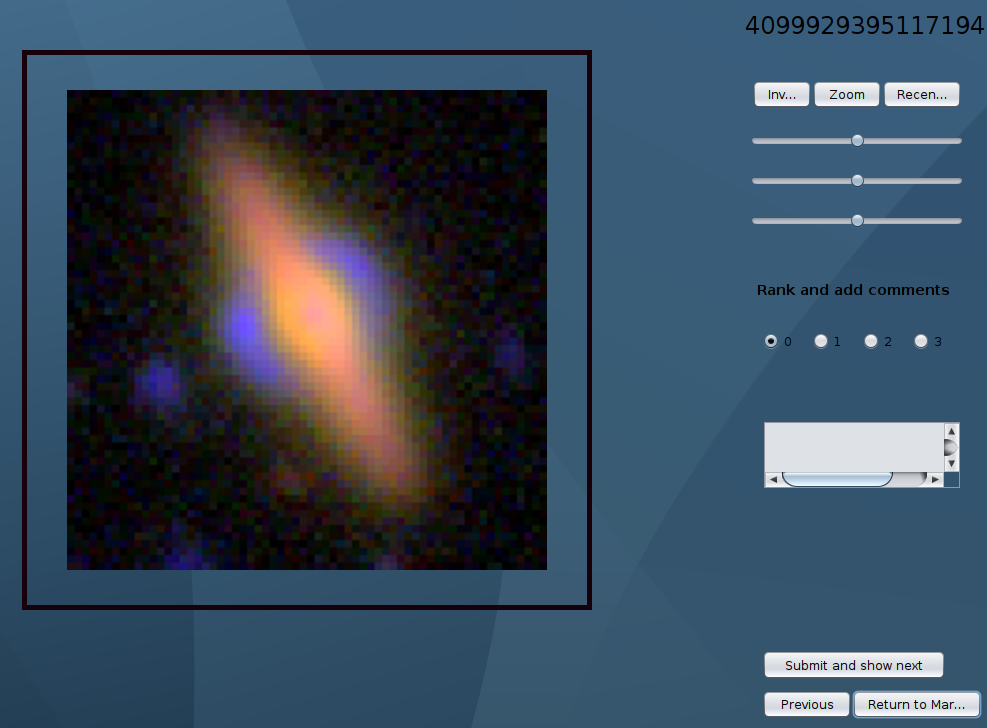}
\caption{Example of an image cutout as viewed on the {\sc visapp}. The cutout is 10.752\arcsec on a side.}
\label{fig:visapp}
\end{figure}

\subsection{Visual Inspection}
\label{sec:vis}

We applied the trained CNN to the $gri$ cutouts of all 2.4 million galaxies in order to determine their score $p_{\rm CNN}$. The CNN identifies 55 469, 20 241, and 1830 as candidate lenses with $p_{\rm CNN}> 0.5$, $p_{\rm CNN}> 0.9$, and $p_{\rm CNN}=1.0$, respectively. About 0.86\% of the parent sample has $p_{\rm CNN}> 0.9$. We note that several candidates have a blue feature close to a red galaxy with a very faint or no counter blue image.  This is possibly due to our training dataset being sensitive to small Einstein radii. After an initial visual inspection performed by a single inspector (ATJ) to remove definite non-lenses, a "good" sample of 1522 potential lens candidates is determined. 

We inspect RGB images of these candidates using {\sc visapp}, which displays PNG images made with $gri$ imaging using three different scaling parameters. \fref{fig:visapp} illustrates a sample page of candidates from the {\sc visapp}. We assign each candidate a grade between 0 and 3, using the grading convention from \citet{Son20}, where 0 means "almost certainly not a lens", 1 means "possibly a lens", 2 means "probably a lens", and 3 means "almost certainly a lens". There are comment columns for the candidates that require further discussion. It is possible to invert the images's colour, zoom in or out and shift it. The grades used throughout the paper are the mean grade, $\langle Gr\rangle$, assigned by authors ATJ, AM, KTI, DCYC, KCW, and PWP. 

\begin{table}
	\centering
	\caption{Results of the lens search on the parent catalogue.}
	\label{tab:result}
	\begin{tabular}{L{5.5cm}C{2.cm}} 
		\hline
		Criteria & Number \\
		\hline
		Image tested  & 2 350 061\\
		First candidates set & \\
		\qquad $p_{\rm CNN}>0.5$    & 55 469\\
		\qquad $p_{\rm CNN}>0.9$    & 20 241\\
		\qquad $p_{\rm CNN}=1.0$    &  1 830\\
		Second candidates set       &  1 522\\
		Final candidates set        &  1 292\\
		\qquad Grade A              &    43\\
		\qquad Grade B              &    269\\
		\qquad\qquad Known (A or B) &    215\\
		\qquad\qquad New (A or B)   &     97\\
        \qquad Grade C              &    880\\
        \qquad\qquad Known (C)      &    289\\
        \hline
        For balanced test sample (50-50) & \\
        \qquad Precision (purity) & \\
        \qquad\qquad ($p_{\rm CNN}>0.5$)  &  97.3\%\\
        \qquad\qquad ($p_{\rm CNN}>0.9$)  &  98.8\%\\
        \qquad Recall (completeness)       & \\
        \qquad\qquad ($p_{\rm CNN}>0.5$)  &  82.3\% \\
        \qquad\qquad ($p_{\rm CNN}>0.9$)  &  72.3\% \\
        For imbalanced test sample (0.3-99.7) & \\
        \qquad Precision (purity) & \\
        \qquad\qquad ($p_{\rm CNN}>0.5$)  &  14.9\%\\
        \qquad\qquad ($p_{\rm CNN}>0.9$)  &  29.3\%\\
        \qquad Recall (completeness)       & \\
        \qquad\qquad ($p_{\rm CNN}>0.5$)  &  82.3\% \\
        \qquad\qquad ($p_{\rm CNN}>0.9$)  &  72.3\% \\
        The fraction of grades A and B with $\langle Gr\rangle>1.5$ & \\
        \qquad $0.95 < p_{\rm CNN}\leq 1.00$ & 268/312\\
        \qquad $0.90 < p_{\rm CNN}\leq 0.95$ &  44/312\\
		\hline
	\end{tabular}
\end{table}

Next, we implemented the scheme outlined below \citep{Son18}: 
\begin{enumerate}
\item[] A: $\langle Gr\rangle>2.5$,
\item[] B: $1.5<\langle Gr\rangle\leq2.5$, 
\item[] C: $0.5<\langle Gr\rangle\leq1.5$, and
\item[] Not a lens: $\langle Gr\rangle\leq0.5$.
\end{enumerate}

Then, the grades were compiled and averaged, and each grader's scores were revealed and compared. Graders were then allowed to adjust their scores, particularly for contentious objects with disparate scores or scores close to the  grade thresholds. During this stage, we also used the HSC PDR2 map\footnote{https://hsc-release.mtk.nao.ac.jp/hscMap-pdr2/app/\#/} for additional inspection, if necessary. 

\begin{figure}
\includegraphics[width=0.47\textwidth]{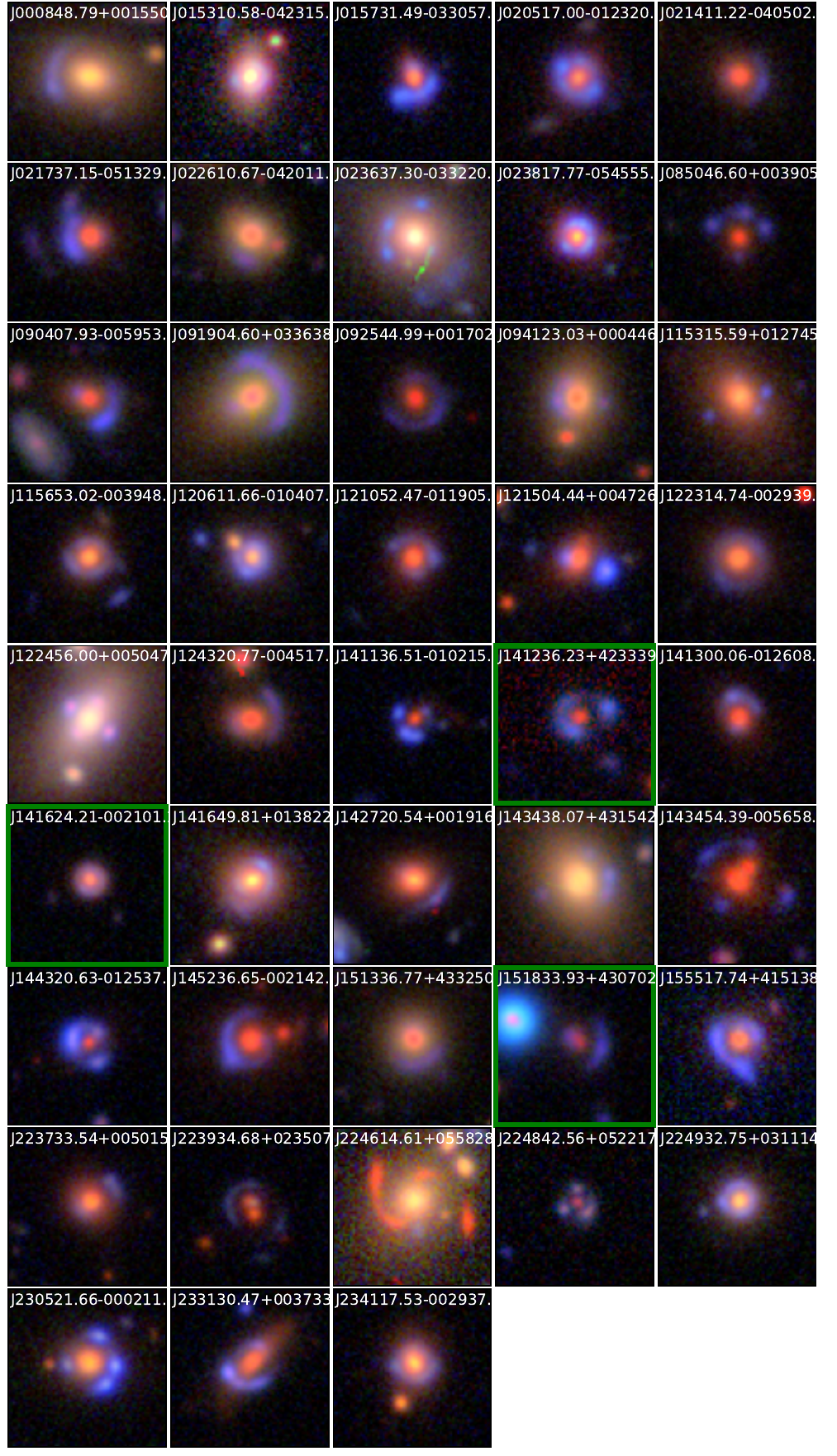}
\caption{Images for the 43 lens candidates with grades A. All images are oriented with North up and East left. Green boxes indicate newly discovered candidates.}
\label{fig:Figure1}
\end{figure}

\begin{table*}
\caption{The grade A and  B strong lens candidates. Column 1 is ID of the system name. Columns 2 and 3 are right ascension and declination of the lens galaxy. Columns 4 and 5 are the \textit{i}-band CModel magnitude and photometric redshift of the lens galaxy provided by the HSC catalogue. Columns 6 gives the spectroscopic redshifts of the lens galaxy inferred from SDSS DR16. Column 7 is the CNN score. Column 8 is the average visual-inspection score of the lens system. 'This Work' indicates a completely new discovery. Other relevant references in column 9 are:(a) \citet{Son18}, (b) \citet{Wong18}, (c) \citet{Son19}, (d) \citet{Chan20}, (e) \citet{Jae20}, (f) \citet{Son20}, (g) \citet{More12}, (h) \citet{More16}, (i) \citet{Pet19}, (j) \citet{Huang20}, (k) \citet{Can20}, (l) \citet{Can21}, (m) \citet{Jac19}, (n) \citet{Cab07}, (o) \citet{Fau08}, (p) \citet{Gav14}, (q) \citet{Li20}, (r) \citet{Bro12}, (s) \citet{Shu16}, (t) \citet{Shu+22}, (u) \citet{Wong+22}. The full list is available at the \href{http://www-utap.phys.s.u-tokyo.ac.jp/~oguri/sugohi/}{http://www-utap.phys.s.u-tokyo.ac.jp/$\sim$oguri/sugohi/}.}
\label{tab:tables}
\begin{tabular}{L{3cm}rrC{1cm}C{1cm}C{1cm}C{1cm}C{1cm}C{3.7cm}}
\hline
HSC\_ID & $\alpha$(J2000) & $\delta$(J2000) &$m_{i, \ell}$ & $\zlp$ & $\zls$& P$_{\rm CNN}$ & $\langle Gr\rangle$ & References \\ 
\hline
J000848.79+001550.8 &   2.20333 &  0.26412 & 19.57 & 0.348 & 0.397 & 0.901 & 2.667 & e, l, t\\	
J015310.58-042315.5	&  28.29412	& -4.38766 & 19.01 & 0.306 &  ...  & 0.990 & 2.667 & t\\	
J015731.49-033057.6 &  29.38125 & -3.51603 & 17.83 & 0.683 & 0.621 & 1.000 & 2.667 & a, l, t\\	
J020517.00-012320.4 &  31.32086 & -1.38901 & 19.57 & 0.770 &  ...  & 0.997 & 2.667 & l, m\\	
J021411.22-040502.8 &  33.54676 & -4.08411 & 17.99 & 0.662 & 0.609 & 0.987 & 2.667 & b, e, g, l, n, p, t\\	
J021737.15-051329.4 &  34.40481 & -5.22486 & 17.96 & 0.629 & 0.646 & 0.996 & 3.000 & c, e, m, n, p, t\\	
J022610.67-042011.6 &  36.54447 & -4.33658 & 18.57 & 0.469 & 0.495 & 0.991 & 2.667 & a, c\\	
J023637.30-033220.1 &  39.15545 & -3.53893 & 20.03 & 0.275 & 0.269 & 1.000 & 3.000 & a, e, l, t\\	
J023817.77-054555.5 &  39.57404 & -5.76543 & 20.13 & 0.594 & 0.599 & 0.999 & 3.000 & a, c, d, l\\	
J085046.60+003905.4 & 132.69420 &  0.65150 & 20.98 & 0.951 &  ...  & 0.988 & 3.000 & f, l, t\\	
J090407.93-005953.0 & 136.03306 & -0.99807 & 18.77 & 0.622 &  ...  & 1.000 & 3.000 & g, l, n, t\\	
J091904.60+033638.5 & 139.76917 &  3.61072 & 20.28 & 0.406 & 0.444 & 0.963 & 2.833 & a, e, k, l, t\\	
J092544.99+001702.8 & 141.43748 &  0.28413 & 20.42 & 0.840 &  ...  & 0.908 & 3.000 & e, f, l, t\\	
J094123.03+000446.6 & 145.34596 &  0.07964 & 21.28 & 0.456 & 0.486 & 1.000 & 2.667 & b, l, t\\	
J115315.59+012745.8 & 178.31498 &  1.46274 & 20.09 & 0.504 &  ...  & 1.000 & 2.833 & e, f, t\\	
J115653.02-003948.5 & 179.22093 & -0.66350 & 21.31 & 0.540 & 0.508 & 0.991 & 2.667 & a, i, t\\	
J120611.66-010407.2 & 181.54861 & -1.06867 & 18.18 & 0.534 &  ...  & 0.997 & 2.833 & l, t\\	
J121052.47-011905.2 & 182.71866 & -1.31814 & 21.26 & 0.683 & 0.700 & 0.998 & 2.667 & a, c, d, l, t\\	
J121504.44+004726.0 & 183.76850 &  0.79057 & 20.32 & 0.655 & 0.642 & 1.000 & 2.667 & c, r\\	
J122314.74-002939.5 & 185.81143 & -0.49432 & 18.26 & 0.541 & 0.547 & 0.990 & 2.667 & b, l, t\\	
J122456.00+005047.9 & 186.23336 &  0.84665 & 17.98 & 0.226 &  ...  & 0.996 & 2.667 & l, q\\	
J124320.77-004517.7 & 190.83655 & -0.75494 & 18.23 & 0.634 & 0.654 & 0.938 & 2.667 & b, i, l, t\\	
J141136.51-010215.6 & 212.90214 & -1.03769 & 20.96 & 0.993 & 0.949& 1.000 & 3.000 & d, f\\	
J141236.23+423339.7 & 213.15097 & 42.56105 & 20.01 & 0.854 &  ...  & 0.996 & 2.667 & This Work\\	
J141300.06-012608.1 & 213.25027 & -1.43561 & 20.81 & 0.757 & 0.749 & 0.999 & 2.667 & a, c, d, l\\	
J141624.21-002101.2 & 214.10091 & -0.35036 & 20.18 & 0.631 &  ...  & 0.981 & 2.667 & This Work\\	
J141649.81+013822.2 & 214.20756 &  1.63951 & 20.02 & 0.313 &  ...  & 1.000 & 2.667 & f, i, l, t\\	
J142720.54+001916.0 & 216.83560 &  0.32111 & 17.32 & 0.558 & 0.551 & 0.981 & 2.833 & a, c, l, t\\	
J143438.07+431542.1 & 218.65864 & 43.26171 & 20.04 & 0.342 & 0.385 & 0.998 & 2.833 & e, l, t\\	
J143454.39-005658.6 & 218.72664 & -0.94964 & 18.36 & 0.729 & 0.728 & 0.948 & 2.667 & a, e, t\\	
J144320.63-012537.2 & 220.83598 & -1.42700 & 18.83 & 0.952 & 0.890 & 0.983 & 2.667 & d, f\\	
J145236.65-002142.1 & 223.15272 & -0.36171 & 18.15 & 0.695 & 0.733 & 0.997 & 2.667 & a, l, t\\	
J151336.77+433250.9 & 228.40322 & 43.54750 & 20.85 & 0.469 & 0.487 & 0.968 & 2.667 & b, l, t\\	
J151833.93+430702.5 & 229.64140	& 43.11739 & 19.90 & 0.986 &  ...  & 0.941 & 2.667 & This Work\\	
J155517.74+415138.6 & 238.82392 & 41.86074 & 20.34 & 0.600 & 0.555 & 1.000 & 2.667 & a, k, l, t\\	
J223733.54+005015.6 & 339.38976 &  0.83767 & 19.97 & 0.634 & 0.605 & 0.986 & 2.667 & a, c, t\\	
J223934.68+023507.4 & 339.89453 &  2.58540 & 17.41 & 0.981 &  ...  & 0.961 & 2.667 & e, f, t\\	
J224614.61+055828.5 & 341.56090 &  5.97459 & 20.00 & 0.274 & 0.340 & 0.957 & 3.000 & e, l, t\\	
J224842.56+052217.7 & 342.17737 &  5.37159 & 20.21 & 0.904 &  ...  & 0.920 & 2.667 & t\\	
J224932.75+031114.4 & 342.38647 &  3.18736 & 20.36 & 0.315 &  ...  & 0.998 & 3.000 & l, t\\	
J230521.66-000211.6 & 346.34029 & -0.03658 & 20.27 & 0.497 & 0.492 & 1.000 & 3.000 & b, l, m, t\\	
J233130.47+003733.4 & 352.87700 &  0.62595 & 20.11 & 0.535 & 0.552 & 1.000 & 2.833 & b, l, m, t\\	
J234117.53-002937.9 & 355.32304 & -0.49388 & 18.99 & 0.511 & 0.531 & 0.979 & 2.833 & l, t, u\\	
\hline
J000326.35-002140.8	&   0.85980 & -0.36133 & 18.34 & 0.794 &  ...  & 1.000 & 1.667 & This Work\\	
J001313.60+002759.9	&   3.30668 &  0.46665 & 18.95 & 0.542 &  ...  & 0.990 & 1.667 & t\\	
J011219.08-001539.3	&  18.07951 & -0.26094 & 20.63 & 0.541 &  ...  & 0.985 & 1.667 & This Work\\	
J011225.76-002247.0 &  18.10737 & -0.37974 & 21.50 & 0.476 & 0.466 & 0.990 & 2.167 & e\\	
J012204.59-002222.1	&  20.51915 & -0.37282 & 20.78 & 1.036 & 0.848 & 0.989 & 1.667 & This Work\\	
J012402.54-004559.3 &  21.01060	& -0.76648 & 20.02 & 0.537 & 0.543 & 0.964 & 1.667 & t\\	
J015702.39-053436.0 &  29.25998	& -5.57669 & 19.39 & 0.319 &  ...  & 0.936 & 2.500 & This Work\\	
J015713.94-045446.9	&  29.30809	& -4.91304 & 21.41 & 0.774 &  ...  & 0.993 & 1.833 & This Work\\	
J020832.14-043315.8 &  32.13393 & -4.55439 & 19.30 & 0.761 &  ...  & 0.988 & 2.000 & e, h, l, t\\	
J095906.48+024524.8 & 149.77704 &  2.75691 & 19.37 & 0.317 & 0.346 & 0.954 & 1.833 & o\\	
J115944.63-000728.2 & 179.93600 & -0.12452 & 20.92 & 0.734 & 0.332 & 0.988 & 2.333 & r\\	
J232557.38-005226.8 & 351.48909 & -0.87412 & 20.11 & 0.794 &  ...  & 0.980 & 2.167 & j, l, m\\	
J234248.68-012032.6 & 355.70284 & -1.34239 & 18.00 & 0.578 & 0.527 & 0.998 & 2.000 & l, s\\	
\multicolumn{1}{c}{\vdots} & \multicolumn{1}{c}{\vdots} & \multicolumn{1}{c}{\vdots} & \multicolumn{1}{c}{\vdots} & \multicolumn{1}{c}{\vdots} & \multicolumn{1}{c}{\vdots} & \multicolumn{1}{c}{\vdots} & \multicolumn{1}{c}{\vdots} & \multicolumn{1}{c}{\vdots}\\
\hline
\end{tabular}
\end{table*}

\section{Results}
\label{sec:res}

\begin{figure}
\includegraphics[width=0.48\textwidth]{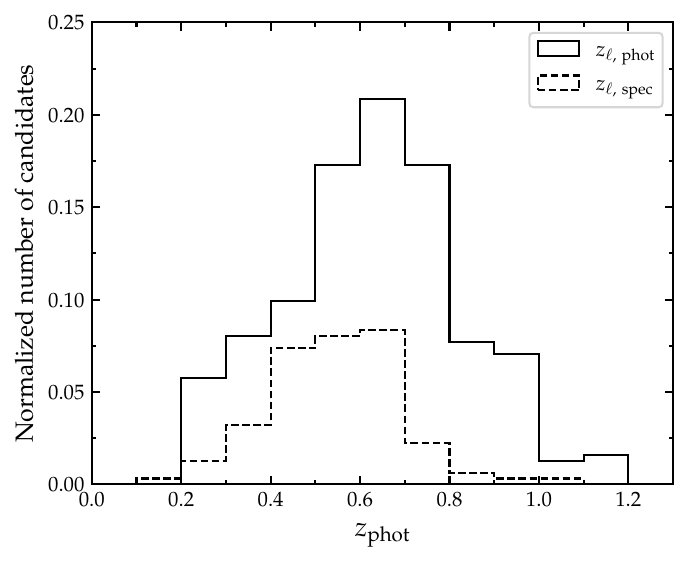}
\caption{Histogram of lens redshifts for the combined grade A and grade B candidates found in this study.}
\label{fig:zphot}
\end{figure}

In total, we discover 43 grade A, 269 grade B, and 880 grade C candidates. The results are summarised in \tref{tab:result}, which include the number of parent sample, lens candidates, purity, and completeness. The full list of the grade A and some examples of grade B candidates are given in \tref{tab:tables}. We note that the nearby lensed arcs could affect the $i$-band magnitude and photometric redshift. We show colour composite images of the grade A candidates in \fref{fig:Figure1} and of the grade B candidates in \fref{fig:Figure2}. The lens redshift distribution of the grade A and grade B candidates is shown in \fref{fig:zphot} with the peak at $z_{\ell,\ \rm{phot}}\sim0.7$ and $z_{\ell,\ \rm{spec}}\sim0.6$. While some lens candidates which were grade C from previous SuGOHI publications have been upgraded to a higher grade in recent references, we list the grade C lens candidates in the Appendix. The fraction of grades A and B with $\langle Gr\rangle>1.5$ among CNN recommendations, decreases rapidly when lowering the threshold $p_{\rm CNN}$. We estimate that 85.9\% of the highest scores $0.95 < p_{\rm CNN}\leq 1.00$ which has $\langle Gr\rangle>1.5$, and about 14.1\% for $0.90 < p_{\rm CNN}\leq 0.95$. In general, we find that the majority of lensed sources have a blue colour, which is due to the selection effect from our training sample.

Separately from this work, \citet{Can21} and \citet{Shu+22} have recently applied machine-learning-based lens search algorithms to the HSC-SSP PDR2 \citep{Aihara+19} and have published lens candidates from these searches. Our parent sample and training set have smaller number of galaxies compared to \citet{Can21} and \citet{Shu+22}, however, we still are able to discover a comparable number of lenses (see details in More et al. in prep). We noted that about 51 percent of our parent sample and the parent sample from \citet{Can21} overlapped. There is some overlap in the candidates we discovered as we used the HSC-SSP dataset. We have identified and cross-matched previous candidates with our candidates and listed the relevant references in \tref{tab:tables}. We find 504 overlapping candidates with 215 and 289 candidates from grade A or B, and grade C, respectively. We find that 143 of 289 overlapping grade C have a higher grade in previous searches. Our CNN also recovers 32 group- and cluster-scale lens candidates, although it is not optimized for these systems. Many candidates with a blue arc feature near a red galaxy have a lower grade, as seen in \tref{fig:Figure3}. \cite{Rojas23} found that is difficult for experts to recover candidates with Einstein radius less than 1.2 times the seeing during visual inspection.

\section{Summary}
\label{sec:summ}

We present new candidates for galaxy-scale strong gravitational lenses from the HSC-SSP PDR2, covering 800 deg$^2$ in the $gri$ bands, using a CNN as part of the SuGOHI project. In order to identify possible galaxy-scale lenses, we use strict pre-selection criteria, such as multiband magnitudes, stellar mass, star formation rate, extendedness limit, and photometric redshift range, resulting in about 2.4 million galaxies of parent sample.  By initially adopting a conventional threshold, $p_{\rm CNN}> 0.9$, we find 20 241 candidates which are then refined to 1 522 systems with lensing features for further inspection. 

On the basis of visual inspections of the cutouts, six authors independently graded the strong lens candidates. According to the average visual inspection scores, 312 candidates identified by the CNN are considered grade A or B (i.e., definite or probable), and 880 are grade C (possible) strong-lens candidates. This list was cross-referenced with our extensive compilation of strong gravitational lenses previously published as confirmed systems or candidates in order to identify duplicates. Among the candidates, there are 97 new discoveries (3 from grade A and 94 from grade B, respectively). We find that 143 grade C candidates were graded as A or B in previous literature. The (photometric) redshift of the candidate lens galaxies range from 0.2 to 1.2. Follow-up spectroscopy will confirm these lenses and measure source redshifts so that detailed lens modelling can yield scientific results.

\section*{Acknowledgements}
We are also grateful to Honoka Murakami and Sherry H. Suyu for their valuable contributions and fruitful discussions, which have significantly improved the quality of this manuscript. This research is supported by Program Riset Unggulan Pusat dan Pusat Penelitian (RU3P) 2023 which is funded by LPIT Insitut Teknologi Bandung. This work also is supported by JSPS KAKENHI Grant Number JP20K14511. R.C. thanks the Max Planck Society for support through the Max Planck Research Group for S.H.S. This project has received funding from the European Research Council (ERC) under the European Union’s Horizon 2020 research and innovation programme (LENSNOVA: grant agreement No 771776).

The Hyper Suprime-Cam (HSC) collaboration includes the astronomical communities of Japan and Taiwan, and Princeton University. The HSC instrumentation and software were developed by the National Astronomical Observatory of Japan (NAOJ), the Kavli Institute for the Physics and Mathematics of the Universe (Kavli IPMU), the University of Tokyo, the High Energy Accelerator Research Organization (KEK), the Academia Sinica Institute for Astronomy and Astrophysics in Taiwan (ASIAA), and Princeton University. Funding was contributed by the FIRST program from the Japanese Cabinet Office, the Ministry of Education, Culture, Sports, Science and Technology (MEXT), the Japan Society for the Promotion of Science (JSPS), Japan Science and Technology Agency (JST), the Toray Science Foundation, NAOJ, Kavli IPMU, KEK, ASIAA, and Princeton University.

This paper is based on data collected at the Subaru Telescope and retrieved from the HSC data archive system, which is operated by Subaru Telescope and Astronomy Data Center (ADC) at NAOJ. Data analysis was in part carried out with the cooperation of Center for Computational Astrophysics (CfCA) at NAOJ. We are honored and grateful for the opportunity of observing the Universe from Maunakea, which has the cultural, historical and natural significance in Hawaii. This paper makes use of software developed for Vera C. Rubin Observatory. We thank the Rubin Observatory for making their code available as free software at \href{http://pipelines.lsst.io/}{http://pipelines.lsst.io/}. Funding for SDSS-III has been provided by the Alfred P. Sloan Foundation, the Participating Institutions, the National Science Foundation, and the U.S. Department of Energy Office of Science. The SDSS-III web site is \href{http://www.sdss3.org/}{http://www.sdss3.org/}. SDSS-III is managed by the Astrophysical Research Consortium for the Participating Institutions of the SDSS-III Collaboration including the University of Arizona, the Brazilian Participation Group, Brookhaven National Laboratory, Carnegie Mellon University, University of Florida, the French Participation Group, the German Participation Group, Harvard University, the Instituto de Astrofisica de Canarias, the Michigan State/Notre Dame/JINA Participation Group, Johns Hopkins University, Lawrence Berkeley National Laboratory, Max Planck Institute for Astrophysics, Max Planck Institute for Extraterrestrial Physics, New Mexico State University, New York University, Ohio State University, Pennsylvania State University, University of Portsmouth, Princeton University, the Spanish Participation Group, University of Tokyo, University of Utah, Vanderbilt University, University of Virginia, University of Washington, and Yale University. 

This research made use of Astropy\footnote{http://www.astropy.org} \citep{astropy13,astropy18}, NumPy \citep{haris16}, matplotlib \citep{caswell19}, {\sc simct} \citep{More16}, and TensorFlow \citep{tensorf22}.

\section*{Data Availability}

The basic information of the lens candidates can be browsed or downloaded from the SuGOHI online database \url{http://www-utap.phys.s.u-tokyo.ac.jp/~oguri/sugohi/} and imaging data from the HSC-SSP Public Data Release 2 \url{https://hsc-release.mtk.nao.ac.jp/doc/}. 


\bibliographystyle{mnras}
\bibliography{sugbib} 

\newpage
\appendix

\section{Images of 269 Grade B Candidate}

\begin{figure*}
\includegraphics[width=0.99\textwidth]{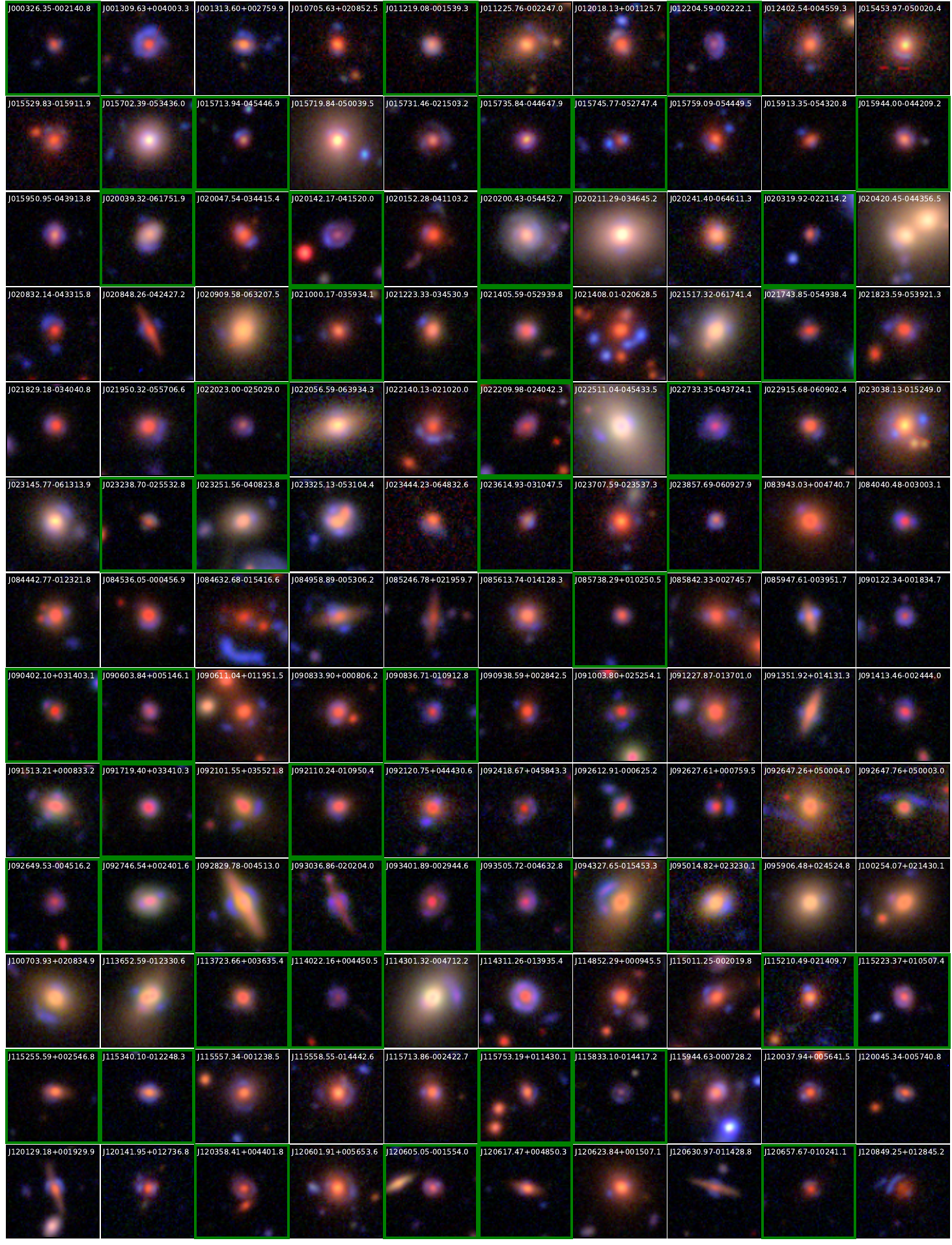}
\caption{Images for the 269 lens candidates with grades B. All images are oriented with North up and East left. Green boxes indicate the new ones.}
\label{fig:Figure2}
\end{figure*}

\begin{figure*}
\ContinuedFloat
\includegraphics[width=0.99\textwidth]{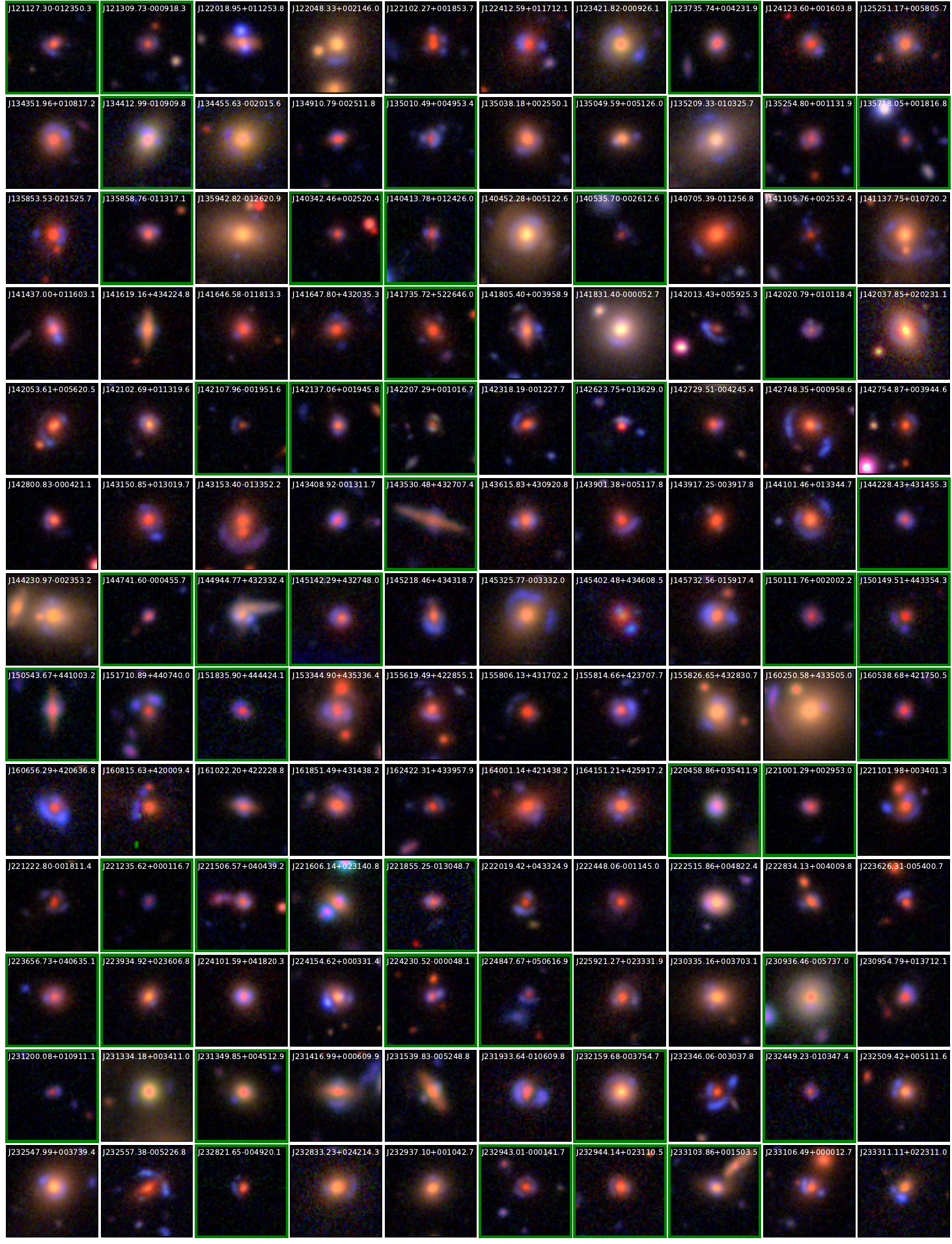}
\caption{\textit{Continued}.}
\end{figure*}

\begin{figure*}
\ContinuedFloat
\includegraphics[width=0.99\textwidth]{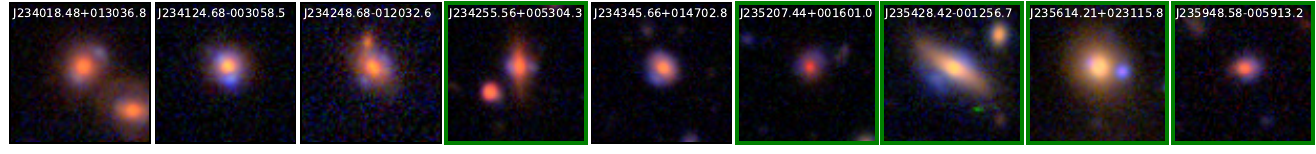}
\caption{\textit{Continued}.}
\end{figure*}

\newpage
\section{Images of 143 Grade C Candidate}

The full list of 880 grade C is also available at the \href{http://www-utap.phys.s.u-tokyo.ac.jp/~oguri/sugohi/}{http://www-utap.phys.s.u-tokyo.ac.jp/$\sim$oguri/sugohi/}. Here we show 143 grade C candidates which had higher grade in previous references.

\begin{figure*}
\includegraphics[width=0.99\textwidth]{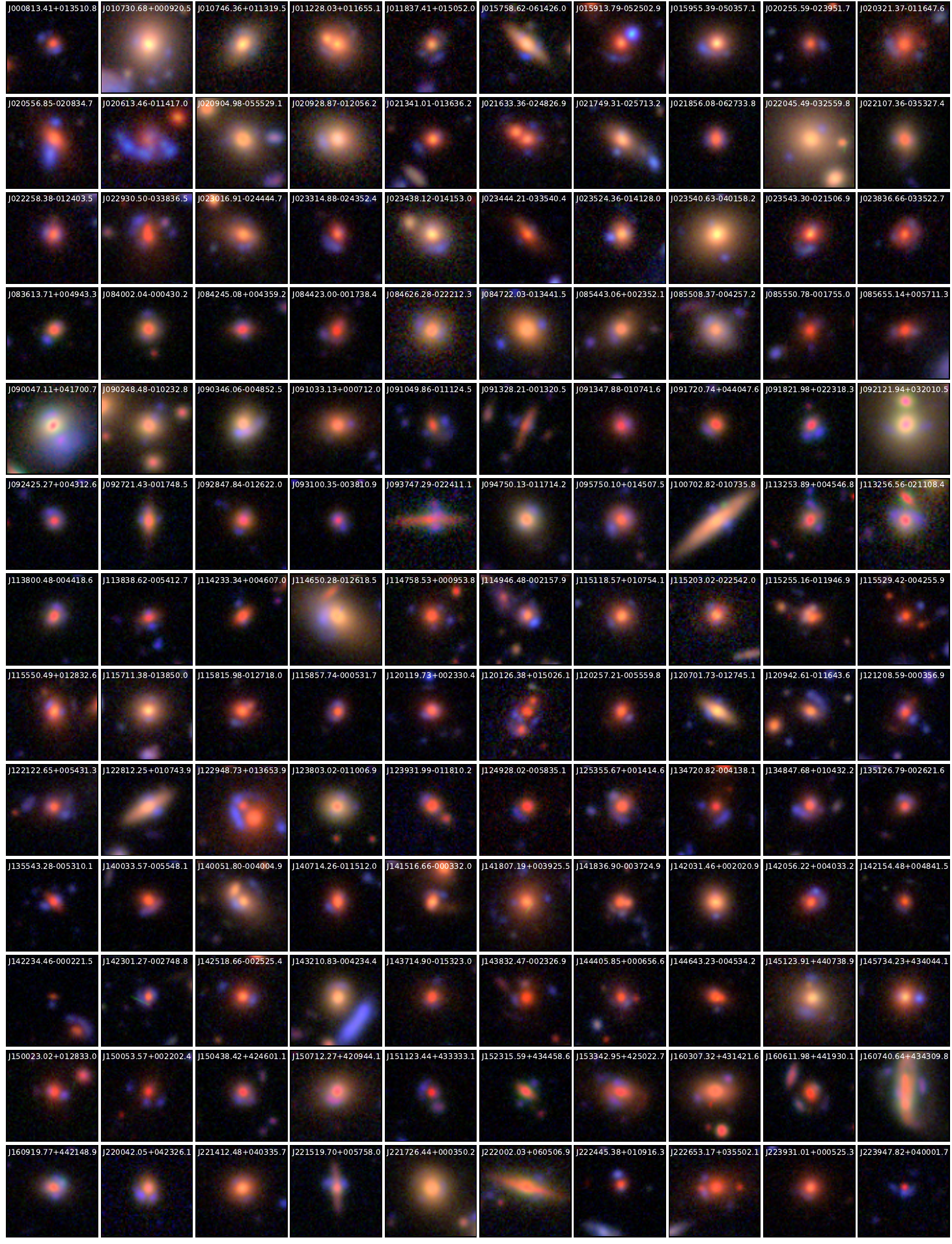}
\caption{Images for the 143 lens candidates with grades C. All images are oriented with North up and East left.}
\label{fig:Figure3}
\end{figure*}

\begin{figure*}
\ContinuedFloat
\includegraphics[width=0.99\textwidth]{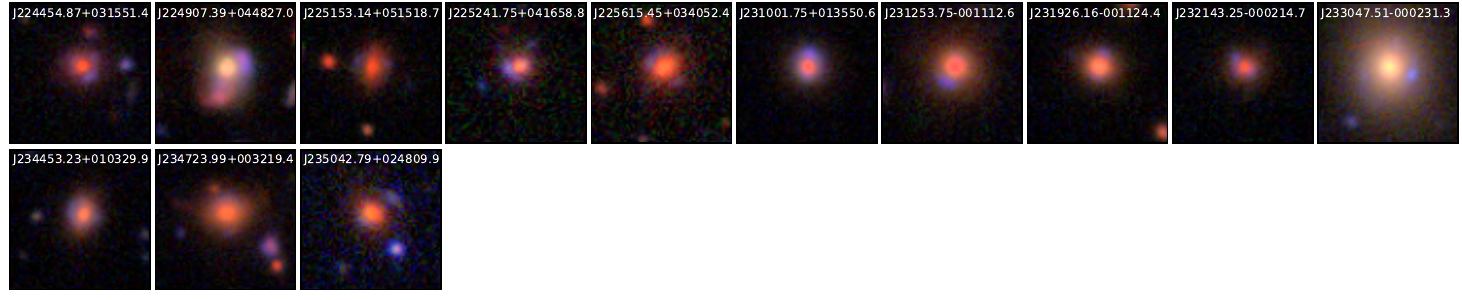}
\caption{\textit{Continued}.}
\end{figure*}


\bsp	
\label{lastpage}
\end{document}